\documentclass[twocolumn,preprintnumbers,amsmath,amssymb,superscriptaddress]{revtex4-2}
\UseRawInputEncoding

\usepackage{latexsym,amssymb,amsthm,amsmath,epsfig,braket}
\usepackage[left=2.00cm, right=2.00cm, top=2.00cm, bottom=2.00cm]{geometry}
\usepackage{xcolor}
\usepackage{svg}
\usepackage[colorlinks, citecolor={blue!80!black}, urlcolor={blue!80!black}, linkcolor={red!50!black}]{hyperref}
\usepackage{float}
\usepackage{hyperref}
\hypersetup{
    citecolor=red,
    colorlinks=true,
    linkcolor=blue,
    filecolor=blue,      
    urlcolor=blue,
}

\begin{document}

\title{Gain-assisted control of the photonic spin Hall effect}

\author{Muhammad Waseem}
\affiliation{Department of Physics and Applied Mathematics, Pakistan Institute of Engineering and Applied Sciences (PIEAS), Nilore $45650$, Islamabad, Pakistan.}
\affiliation{Center for Mathematical Sciences, PIEAS, Nilore, Islamabad $45650$, Pakistan.}
\author{Muzamil Shah}
\email{muzamill@zjnu.edu.cn}
\affiliation{Department of Physics, Zhejiang Normal University, Jinhua, Zhejiang 321004, China}
\affiliation{Zhejiang Institute of Photoelectronics and Zhejiang Institute for Advanced Light Source, Zhejiang Normal University, Jinhua, Zhejiang 321004, China}
\author{Gao Xianlong}
\affiliation{Department of Physics, Zhejiang Normal University, Jinhua, Zhejiang 321004, China}

\date{\today}

\begin{abstract}
In the photonic spin Hall effect (SHE), also known as the transverse shift, incident light photons with opposite
spins are spatially separated in the transverse direction due to the spin-orbit interaction of light. Here, we propose
a gain-assisted model to control the SHE in the reflected probe light. In this model, a probe light is incident on
a cavity containing a three-level dilute gaseous atomic medium, where the interaction between the atom and the
control field follows two-photon Raman transitions. We show that the direction of photonic spin accumulations
can be switched between positive and negative values across the Brewster angle in both the anomalous and
normal dispersion regimes. For the same magnitude of control fields, the peak value of the photonic SHE is
higher in the anomalous dispersion region compared to the normal dispersion regime. Additionally, the angular
range around the Brewster angle is wider in the normal dispersion regime than in the anomalous dispersion
region. Furthermore, the peak value of the photonic SHE and the angular range can be controlled by changing
the Rabi frequencies of the control fields and the probe field detuning. The measurement of photonic SHE based
on gain assistance may enable spin-related applications such as optical sensing.
\end{abstract}
\maketitle
\newpage
\section{Introduction}
The photonic spin Hall effect (SHE) is an optical phenomenon
in which photons with opposite spins are separated
from each other due to the spin-orbit interaction of light~\cite{onoda_Hall_2004,kim_spin_2023}.
As a result of spin-dependent splitting, a linearly polarized
beam, composed of right and left circular polarization components,
shifts transversely to the incident plane. The SHE of
light can be regarded as a direct optical analogy of the SHE
in an electronic system. In photonic SHE, the spin and refractive
index gradient of the photons play a role similar to the
spin and electric potential of the electrons, respectively~\cite{bliokh_conservation_2006, hosten_observation_2008}.        
This phenomenon can be traced back to the Imbert-Fedorov
displacement~\cite{fedorov1955polarization, imbert_calculation_1972} and is nowadays referred to as transverse
shift. Recently, the photonic SHE has been intensively investigated
in different physical systems. For example, in a
static gravitational field~\cite{gosselin_spin_2007}, inside semiconductors via absorption~\cite{menard_imaging_2009}, graphene layers~\cite{zhou_identifying_2012,PhysRevA.95.013809,SHAH2024107676}, surface plasmon resonance systems~\cite{salasnich_enhancement_2012,zhou_enhanced_2016,tan_enhancing_2016,xiang_enhanced_2017,wan_controlling_2020}, metamaterials~\cite{yin_photonic_2013}, all-dielectric metasurfaces~\cite{kim_reaching_2022}, topological insulators \cite{SHAH2022115113}, and two dimensional quantum materials \cite{Shah_2022,kort-kamp_topological_2017}. 
These studies highlight the SHE’s
unique properties and intriguing applications, such as probing
topological phase transitions~\cite{Shah_2022, kort-kamp_topological_2017}, identifying graphene layers~\cite{zhou_identifying_2012}, chiral molecular detection~\cite{tang_optimal_2023}, and performing mathematical operations and edge detection~\cite{zhu_generalized_2019}. 
However, the photonic SHE typically appears on a nanometer scale, making direct measurement challenging.
Recently, a large
photonic SHE and high efficiency for arbitrarily incident
polarized light is proposed by exploiting total external and
internal reflection~\cite{kim_total_2021}, and by using anisotropic impedance
mismatching in the microwave spectrum~\cite{kim_spin1_2021}. 
Experimental
results using quantum weak measurements~\cite{kim_nanophotonic-assisted_2022} showed that
the SHE of light can be enhanced near the Brewster angle on
reflection~\cite{qin_measurement_2009,luo_enhanced_2011}.

Optical SHE is another phenomena similar to the photonic
SHE~\cite{kavokin_optical_2005}.
Although both effects involve spin separation, the
exact configurations or origins are not same. The optical
SHE describes optically generated spin currents of excitonpolaritons
in semiconductor microcavities~\cite{leyder_observation_2007, lafont_controlling_2017}. In other words, the term optical SHE is typically used for the spin separation of optically induced polaritons, that scatter in the plane of the microcavity~\cite{lekenta_tunable_2018}. For example,  Leyder et al.,~\cite{leyder_observation_2007} focused linearly polarized pumped photons on microcavities and observed photons with opposite circular polarization scattered in different directions. This methodology is similar to that is used in the photonic SHE literature.
The direction of
the spin currents in the microcavity can be altered by rotating
the polarization plane of the exciting light~\cite{kavokin_optical_2005}.


\begin{figure*}[t]
\begin{tabular}{@{}cccc@{}}
\includegraphics[width=5.25 in]{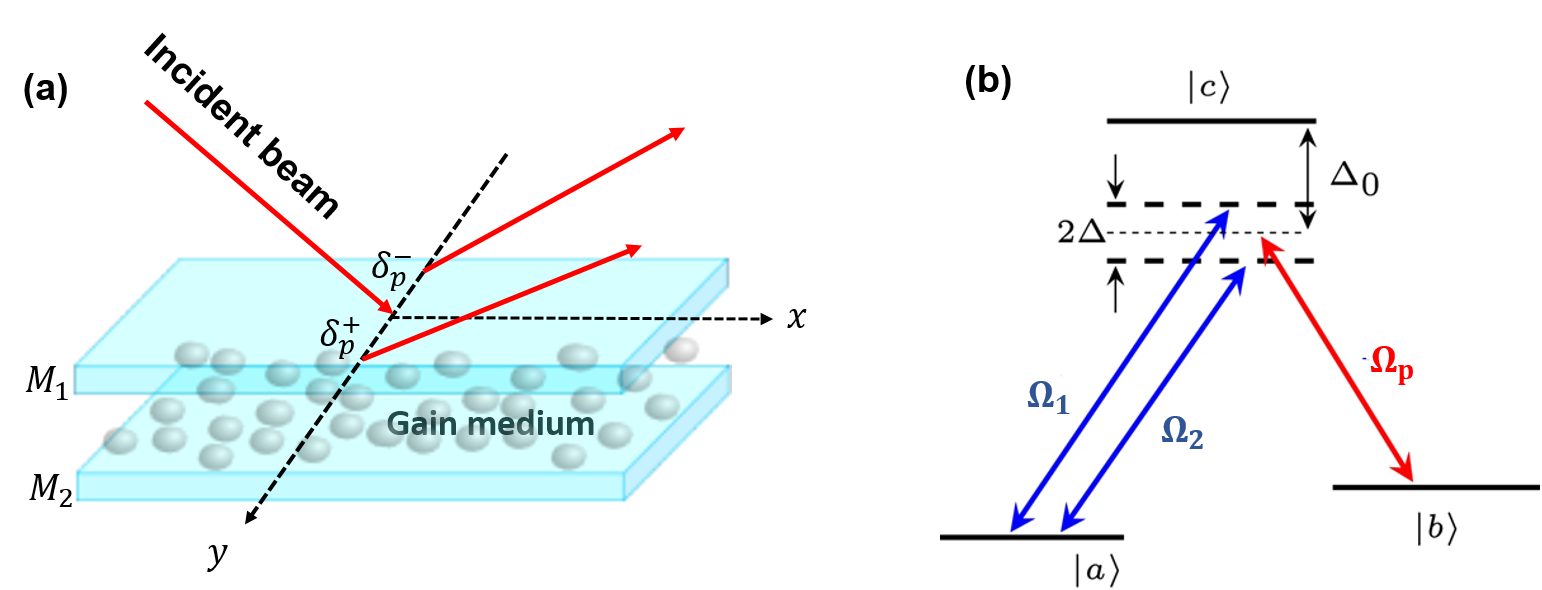}
\end{tabular}
\caption{(a) Schematic setup of a three-layer cavity system composed of two mirrors $M_1$ and $M_2$ containing a coherent Raman gain atomic
medium in between them. Spin-dependent splitting occurs for a TM-polarized incident light reflected on mirror surface $M_1$. (b) Energy level
diagram of three-level Raman gains atomic medium driven by three lasers: probe field and two driving fields.}
\label{fig:sys}
\end{figure*}

However, in cavity quantum electrodynamics (QED), the
interaction between light and atoms induces atomic coherence
and quantum interference effects. These coherence and quantum
interference effects led to many important phenomena,
such as amplification without inversion~\cite{scully_degenerate_1989}, electromagnetically induced transparency~\cite{boller_observation_1991}, slow and fast light~\cite{schmidt_steep_1996}, quantum memory~\cite{kash_ultraslow_1999}, optical solitons~\cite{wu_ultraslow_2004} and so forth.
Another notable example is to manipulate the lateral shift
(shift parallel to the incident plan) known as the Goos-
Hanchen (GH) effect of a light beam reflected or refracted
from the interface of different media and structures~\cite{ziauddin_coherent_2010,wang_control_2008,ziauddin_giant_2015,zhang_Goos-Hanchen_2015,shui_squeezing-induced_2019,asiri_controlling_2023}. 
The GH shift is a result of the spatial dispersion of either
transverse magnetic (TM) or transverse electric (TE) beam
reflection coefficients and the interference of the angular spectrum
components (in other words phase change of reflection
coefficient)~\cite{wang_control_2008,asiri_controlling_2023,ziauddin_gain-assisted_2011}.
The magnitude of the GH shift can be
positive and negative depending on the dispersive properties
of the cavity~\cite{ziauddin_coherent_2010}.
In past extensive research was carried out
to explore GH shift in cavity QED. However, photonic SHE
in cavity QED has not been explored in detail, which is the
subject of present work.

In this paper, we study photonic SHE using atomic coherence
and quantum interference in cavity QED. It was
already established that the dispersion absorption properties
of an atomic medium can be modified using coherent control
of the driving fields~\cite{wang_gain-assisted_2000}.
level atomic medium has been
used via two-photon Raman transitions to obtain gain-assisted
normal and anomalous light propagation~\cite{wang_gain-assisted_2000,dogariu_transparent_2001} and GH shift~\cite{ziauddin_gain-assisted_2011, asiri_controlling_2023}.
Following a similar scheme, we here study the
behavior of the photonic SHE for the corresponding normal
and also for the anomalous propagation of the light through
the intracavity gain-assisted medium. Our results indicate that
the direction of the photonic SHE can be switched between
positive and negative values across the Brewster angle in
both the anomalous and normal dispersion regions. We find
that the peak value of the photonic SHE can be externally
controlled by adjusting the Rabi frequencies of the control
fields and the detuning of the probe field, all without modifying
the physical structure of the material. This flexible
control of the peak value around the Brewster angle, in
combination with weak measurement techniques~\cite{qin_measurement_2009}, may offer a promising approach in cavity quantum electrodynamic
systems.

The rest of the paper is organized as follows. Section II
describes the physical system under consideration and the
corresponding mathematical expressions. Section III provides
details of the results, plots, and discussions regarding the
observed particular behaviors. Finally, the entire discussion
of these interesting findings is summarized in Sec. IV.

\section{Theoretical Model}
Our proposed model is a cavity, which is composed of
three layers as shown in Fig.~\ref{fig:sys}(a). Layers 1 and 3 are the
two dielectric mirrors $M_1$ and $M_2$ while layer 2 is the intracavity
three-level Raman gain medium trapped inside the
two mirrors. In general, mirrors can also be replaced by a
prism~\cite{asiri_controlling_2023}. The thickness of each dielectric mirror is $d_1$ and the intracavity atomic gas medium is $d_2$. The permittivities of
each mirror are $\epsilon_1=2.2$, whereas the permittivity of atomic
medium is $\epsilon_2$. The permittivity of the intracavity medium can
be defined in terms of its dielectric susceptibility as~\cite{wang_control_2008} 
\begin{equation}
\epsilon_{2} = 1 + \chi,
\end{equation}
where $\chi$ is a complex quantity, represents the dielectric susceptibility of the intracavity Raman gain medium, and can be expressed as
\begin{equation}
\chi = \chi_{1} +i \chi_{2},
\end{equation}
where, real part $\chi_{1}$ represents dispersion and imaginary part $\chi_{2}$ represents the absorption of the probe field.

To obtain the dielectric susceptibility of the intracavity
medium, we consider the, $\Lambda$-type atomic system having an excited state $|c\rangle$ and two ground states $|a\rangle$ and $|b\rangle$ as shown in Fig. \ref{fig:sys}(b). 
Two control laser fields of amplitude $E_1$ and $E_2$  with frequencies $\nu_{1}$ and $\nu_{2}$ interacts for off-resonantly with atomic transition $|a\rangle  \leftrightarrow  |c\rangle$.
Initially, the atoms are prepared in state $|a\rangle$ using optical pumping.
The probe field is applied across the transition $|b\rangle  \leftrightarrow  |c\rangle$ with amplitude $E_p$ and frequency $\nu_p$. The energy levels $|b\rangle$ and $|c\rangle$ are not populated.
The effective Hamiltonian of the system can be written as 
\begin{equation}
H=H_0+H_1,
\end{equation}
where the unperturbed part of the Hamiltonian is
\begin{equation}
H_0=-\hbar \omega_{c a}|a\rangle\left\langle a\left|-\hbar \omega_{c b}\right| b\right\rangle\langle b|.
\end{equation}
Here, $\omega_{c a}=2 \pi v_{c a}$ and $\omega_{c b}=2 \pi v_{c b}$ are the angular frequencies corresponding to the respective atomic transitions. 
The perturbed part of the Hamiltonian is 
\begin{eqnarray}
\label{E6}
H_{1} &=& -\hbar (\Omega_1 e^{-i \omega_1 t}+\Omega_2 e^{-i \omega_2 t})\ket{c}\bra{a} \\ \nonumber
&-& \hbar \Omega_p e^{-i \omega_p t} \ket{c} \bra{b}-H.c,
\end{eqnarray}
Here, $\omega_1=2 \pi v_1, \omega_2=2 \pi v_2$, and $\omega_p=2 \pi v_p$ are the angular frequencies associated with the driving laser field $E_1, E_2$ and the probe field $E_p$, respectively. The corresponding Rabi frequencies are $\Omega_1=\mu_{c a}\left|E_1\right| / \hbar, \Omega_2=\mu_{c a}\left|E_2\right| / \hbar$, and $\Omega_p=$ $\mu_{c b}\left|E_p\right| / \hbar$. We consider that the driving laser fields $E_1$ and $E_2$ are strong fields while the probe field $E_p$ is a weak field, which means $\left|\Omega_1\right|\left|\Omega_2\right| \gg\left|\Omega_p\right|$. Using the definition of dielectric polarization and following the density matrix approach~\cite{scully1997quantum}, the dielectric susceptibility for the Raman gain process can be written as
\begin{equation}
\label{eq:chi}
\chi=\frac{M_1}{\left(\delta_p-\Delta v\right)+i \Gamma}+\frac{M_2}{\left(\delta_p+\Delta v\right)+i \Gamma},
\end{equation}
where $\delta_p=v_p-v_0$ is the probe field detuning and $v_0=v_1-$ $v_{c a}+v_{c b}$. The Raman transition inverse lifetime is $\Gamma$, which represents the spectral width of two gain lines. The driving fields $E_1$ and $E_2$ are detuned from each other by an amount $2 \Delta v$, where $\Delta v$ is the difference between $v_1$ and $v_1$. We define
\begin{equation}
M_j=N \frac{\left|\mu_{c b}\right|^2}{4 \pi \hbar \epsilon_0} \frac{\left|\Omega_j\right|^2}{\Delta_0^2},
\end{equation}
where $j=1,2, N$ is the number density of the atom, $\mu_{c b}$ is the dipole matrix element of the transition $c \leftrightarrow b$, and $\epsilon_0$ is the permittivity of vacuum. The common detuning is $\Delta_0=\left(\Delta_1+\right.$ $\left.\Delta_p\right) / 2$ with $\Delta_1=2 \pi\left(v_1-v_{c a}\right)$ and $\Delta_p=2 \pi\left(v_p-v_{c b}\right)$. We consider the condition that $\Delta_0 \gg \Delta_1-\Delta_p$. It is clear from Eq.~\eqref{eq:chi} that the susceptibility of the Raman gains medium and hence its permittivity $\epsilon_2$ can be modified and controlled by changing several parameters such as the Rabi frequencies of the driving fields $\Omega_1$ and $\Omega_2$ and the probe field detuning $\delta_p$. We consider that a TE- and TM-polarized probe light beam is incident on the cavity mirror $M_1$ from the vacuum with an angle of incidence $\theta_i$. This monochromatic Gaussian probe beam will be reflected at the structure interface or pass through the structure. After reflection, the left and right circular polarization components of the incident light beam will be spatially separated in the direction perpendicular to the plane of incidence ( $y$ axis) as shown in Fig.~\ref{fig:sys}(a). This perpendicular shift is known as the photonic SHE. The photonic SHE is inherently a polarization-dependent optical phenomenon of incident light in which photons with opposite helicity are separated from each other due to the spin-orbit interaction of light. The photonic SHE can be considered as an optical version of the electron SHE~\cite{bliokh_spinorbit_2015}. For the three-layer structure discussed here, the complex reflection coefficients for TM-polarized $r_p$ and TE-polarized $r_s$ can be written as~\cite{wan_controlling_2020}
\begin{figure}[t]
\begin{tabular}{@{}cccc@{}}
\includegraphics[width=3.25 in]{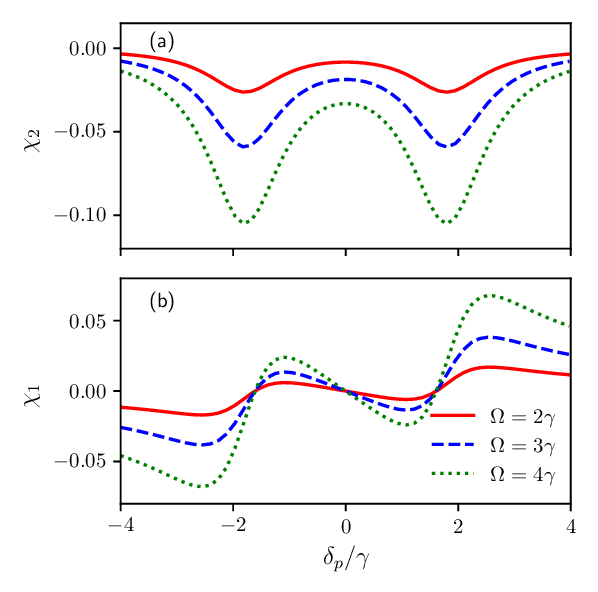}
\end{tabular}
\caption{(a) Absorption spectrum and (b) Dispersion spectrum of the weak probe field as a function of probe detuning $\delta_{p} / \gamma$ at different values of control fields $\Omega=\Omega_{1}=\Omega_{2}$. The region at resonance $\delta_{p}=0$ is called the anomalous dispersion regime, while the region at $\delta_p = \Delta\nu = 1.8 \gamma$ is called the normal dispersion region. The others parameters are $\Gamma = 0.8 \gamma$, $\Delta_1 = 5 \gamma$, $\Delta_p = 4.9 \gamma$, $\mu_{cb} = 3.79 \times 10^{-29}$, $\epsilon_{0} = 8.85 \times 10^{-12}$ F/m, $N = 10^{12}/\text{cm}^3$, and $\lambda = 852$ nm, where $\gamma = 10^6$~Hz. }
\label{fig:chi}
\end{figure}
\begin{figure*}[t]
\begin{tabular}{@{}cccc@{}}
\includegraphics[width=6.35 in]{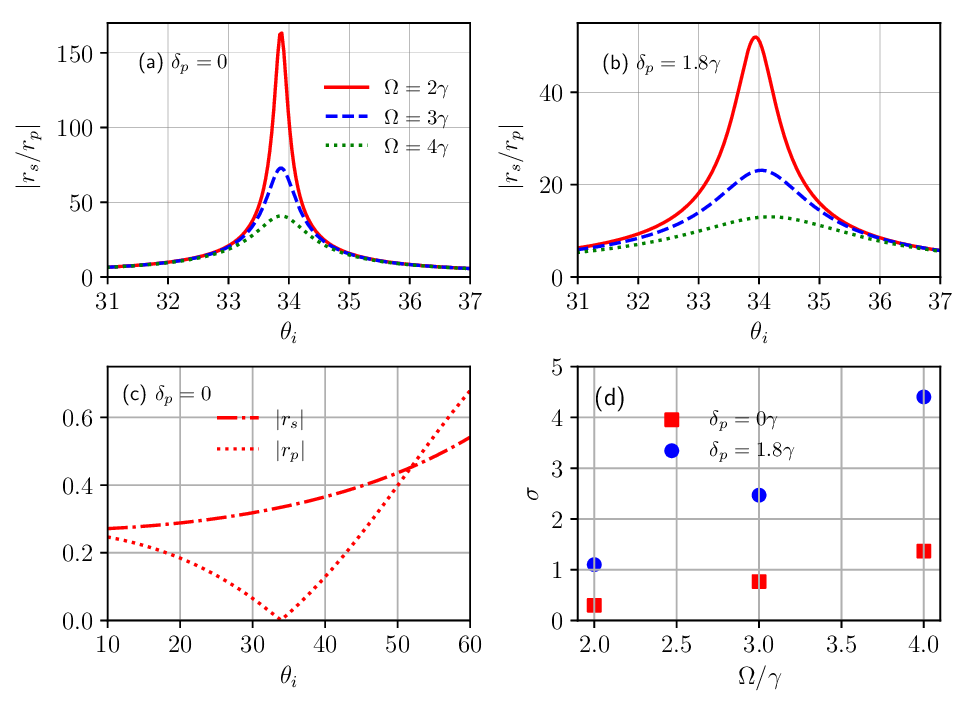}
\end{tabular}
\caption{Ratio of $\left|r_s / r_p\right|$ as a function of incident angle $\theta_i$ in the (a) anomalous dispersion region at $\delta_p=0$ and (b) normal dispersion region at $\delta_p=1.8 \gamma$ at three different values of $\Omega=2 \gamma$ (solid), $\Omega=$ $3 \gamma$ (dashed), and $\Omega=4 \gamma$ (dotted). (c) Plot of individual $\left|r_s\right|$ and $r_p \mid$ as a function of incident angle $\theta_i$ in the anomalous dispersion region at $\delta_p=0$. At Brewster angle $\theta_b$ reflection coefficient $r_p \mid$ converge to zero, which results the enhancement of ratio $\left|r_s / r_p\right|$. (d) Full-width at half maximum $\sigma$ as a function of $\Omega$ in anomalous dispersion region (squares) and normal dispersion region (circles). We consider the interaction length of the atomic medium $d=100 \mathrm{~nm}$. The dielectric permittivities of two mirrors are $\epsilon_1=2.22, \epsilon_3=2.22$. The rest of the parameters are the same as for Fig.~\ref{fig:chi}.}
\label{fig:rsp}
\end{figure*}
\begin{equation}
\label{rp}
r_{p, s}=\frac{r_{p, s}^{12}+r_{p, s}^{23} e^{2 i k_{2 z} d}}{1+r_{p, s}^{12} r_{p, s}^{23} e^{2 i k_{2 z} d}},
\end{equation}
where $r_{p, s}^{i j}$ is the Fresnel's reflection coefficient at the $i-j$ interface given by
\begin{equation*}
r_p^{i j}=\frac{k_{i z} / \varepsilon_i-k_{j z} / \varepsilon_j}{k_{i z} / \varepsilon_i+k_{j z} / \varepsilon_j}, \\
r_s^{i j}=\frac{k_{i z}-k_{j z}}{k_{i z}+k_{j z}} .
\end{equation*}
Here $k_{i z}=\sqrt{k_0^2 \varepsilon_i-k_x^2}$ represents the normal wave vector in the corresponding layer, and $k_x=\sqrt{\varepsilon_1} k_0 \sin \theta_{i}$ is the wave vector along the $x$ direction. 
Here, $k_0=2 \pi / \lambda$ denotes the wave vector with $\lambda$ being the light wavelength.
It can be seen from Eq.~\eqref{rp} that the reflection coefficients depend on the gain medium permittivity $\epsilon_2$, which can be effectively controlled by manipulating $\chi$. This leads to a controllable photonic SHE of light.
Having calculated the dielectric susceptibility for the Raman gain process, we can calculate the photonic SHE in terms of Fresnel’s reflection coefficients. 
The corresponding transverse spin-displacements $\delta_p^{+}$ and $\delta_p^{-}$ can be expressed in terms of the reflective coefficients of the three-layer atomic system~\cite{xiang_enhanced_2017,tan_enhancing_2016}:
\begin{equation}\label{eq:shift}
\delta_p^{ \pm}=\mp \frac{k_1 w_0^2 Re\left [1+\frac{r_s}{r_p} \right]  \cot \theta_{i}}{k_1^2 w_0^2+\left|\frac{\partial \ln r_p}{\partial \theta_{i}}\right|^2+\left|\left(1+\frac{r_s}{r_p}\right) \cot \theta_{i}\right|^2} .
\end{equation}
Here $k_1=\sqrt{\varepsilon_1} k_0$ and $w_0$ represents the radius of the waist of the incident beam. 

\begin{figure*}[t]
\begin{tabular}{@{}cccc@{}}
\includegraphics[width=6.25 in]{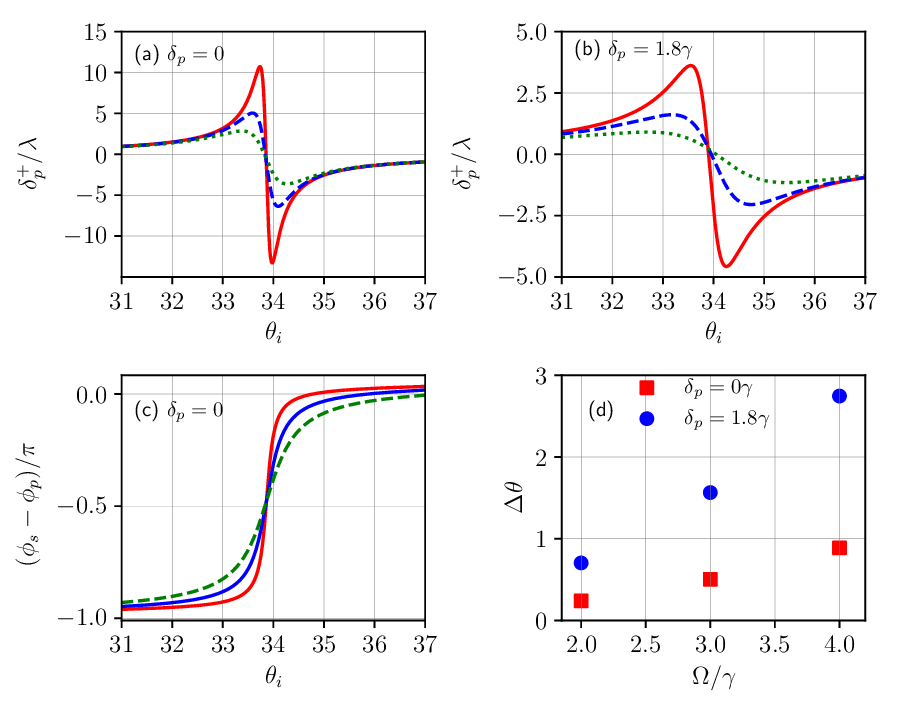}
\end{tabular}
\caption{(a) Photonic spin Hall shift $\delta_{p}^{+}$ as a function of incident angle  $\theta_i$ in the (a) anomalous dispersion region at $\delta_{p}=0$ and (b) normal dispersion region at $\delta_{p}=1.8 \gamma$ at three different values of $\Omega=2 \gamma$ (solid), $\Omega=3 \gamma$ (dashed), and $\Omega=4 \gamma$ (dotted).
(c) The phase difference $\phi_s - \phi_p$ in the anomalous dispersion region experiences a $\pi$ phase variation across the Brewster angle, and the spin accumulation reverses its directions accordingly.
(d) Angular width $\Delta \theta = \theta^{-} - \theta^{+}$ of incident angle as a function of $\Omega$ in anomalous dispersion region (squares) and normal dispersion region (circles). Angular width $\Delta \theta$ shows the range of incident angle over which photonic spin Hall shift changes sign from positive to negative.
The rest of the parameters are the same as for Fig.~\ref{fig:rsp}}
\label{fig:shift}
\end{figure*}

\begin{figure*}[t]
\begin{tabular}{@{}cccc@{}}
\includegraphics[width=5.8 in]{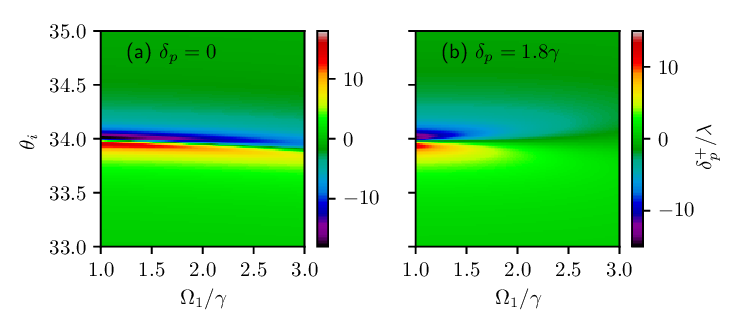}
\end{tabular}
\caption{(a) Contour plot of the photonic Hall shift as a function of incident angle $\theta_i$ and $\Omega_1$ in anomalous dispersion regime $\delta_p=0$ at fixed $\Omega_2=2.0 \gamma$ (b) Contour plot of the photonic Hall shift as a function of incident angle $\theta_i$ and $\Omega_1$ in normal dispersion regime $\delta_p=\Delta \nu$ at fixed $\Omega_2=2.0 \gamma$. The rest of the parameters are unchanged.}
\label{fig:omega}
\end{figure*}

\section{Results and discussion}
Based on the fact that the optical properties of the Raman medium can be controlled, we now discuss the results of the photonic SHE of the reflected probe beam, which can be modified and controlled. To do so, we first consider the susceptibility analysis with $\Omega_1=\Omega_2=\Omega$ which consequently gives $M_1=M_2$. We choose the parameters related to Cs vapors used in experiment~\cite{wang_gain-assisted_2000}. These parameters are $\Gamma=0.8 \gamma, \Delta v=1.8 \gamma, \Delta_1=5 \gamma, \Delta_p=4.9 \gamma, \mu_{c b}=$ $3.79 \times 10^{-29} \mathrm{Cm}, \epsilon_0=8.85 \times 10^{-12} \mathrm{Fm}^{-1}, N=10^{12} / \mathrm{cm}^3$, and $\lambda=852 \mathrm{~nm}$, where $\gamma=10^6 \mathrm{~Hz}$. The interaction length of the atomic medium is $d=100 \mathrm{~nm}$. The corresponding dielectric permittivities of the two mirrors are $\epsilon_1=2.22, \epsilon_3=2.22$. We first plot the imaginary and real parts of $\chi$ as a function of probe detuning $\delta_p$ in Fig.~\ref{fig:chi}(a) and (b), respectively. The imaginary part shows the absorption, while the real component exhibits the dispersion of the probe field. It must be noted that when the incident probe field interacts resonantly with the Raman gain medium, the medium experiences zero dispersion and a small amount of gain. The amount of gain increases as the strength of $\Omega$ increases while dispersion does not depend on it. This region is known as the anomalous dispersion region. Slightly away from resonance and between the two Raman gain doublets, the probe field dispersion becomes negative. At detuning $\Delta \nu =1.8 \gamma$, the dispersion again becomes zero with a larger gain and this region is known as the normal dispersion region. In this way, the optical response of the Raman gain medium can be obviously controlled and this is the motivation to use this system to coherently control the photonic SHE in this work.

It is evident from the Eq.~\eqref{eq:shift}, transversal shift $\delta_p$ depends on the ratio $r_s$ to $r_p$ for an arbitrary incident angle. Its value larger than one can give rise to a noticeable transverse shift.
Therefore, we plot the ratio of $|r_s / r_p|$ as a function of the incident angle $\theta_i$ in the anomalous dispersion region at $\delta_{p}=0$ shown in Fig.~\ref{fig:rsp} (a) and the normal dispersion region at $\delta_{p}=1.8 \gamma$ shown in Fig.~\ref{fig:rsp}(b). The solid, dashed, and dotted curves correspond to $\Omega=2 \gamma$, $\Omega=3 \gamma$, and $\Omega=4 \gamma$, respectively.
The peak value of $|r_s / r_p|$ for the anomalous dispersion region is three times higher than the normal dispersion region.
The ratio increases significantly in the vicinity of the Brewster angle $\theta_{B} \approx 33.9^\circ$ for both normal and anomalous dispersion regions. The reason is that $|r_p|$ response vanishes at Brewster angle as shown by the dotted curve in Fig~\ref{fig:rsp}(c) for the case of $\delta_{p}=0$. This disappearing response of $\left|r_p\right|$ results in an improvement of $\left|r_s / r_p\right|$. Here, we choose the narrow range of incident angle $\theta_i$ for the sake of brewster angle clarity. From the spectrum of $\left|r_s / r_p\right|$ shown in Fig. 3 we extract the full width at half maximum $\sigma$, which quantifies the angular range of $\theta_i \in\left[0^{\circ}, 89^{\circ}\right]$ over which $\left|r_s / r_p\right| \gg 1$. In Fig~\ref{fig:rsp}(d), the squares and circles show the values of $\sigma$ versus $\Omega$ for anomalous and normal dispersion, respectively. Clearly angular range increased with the increase of $\Omega$ and became wider for the normal dispersion regime compared to the normal dispersion regime. In other words, the result in Fig.~\ref{fig:rsp}(d) validates that $\left|r_s / r_p\right| \gg 1$ can be achieved in a very wide angular range by exploiting the anomalous dispersion regime. These results also indicate that when the angular range is wider, the maximum value of $\left|r_s / r_p\right|$ decreases and vice versa.

Next, we analyze the transverse shift due to the photonic SHE. We only present the transverse shift of the right circularly polarized photon spin-dependent component $\delta_p^{+}$because the beam shifts for the two circular components are equal in magnitude and opposite in sign. We plot the photonic spin Hall shift $\delta_p^{+} / \lambda$ as a function of the incident angle $\theta_i$ in the anomalous dispersion region at $\delta_p=0$ shown in Fig. 4(a) and normal dispersion region at $\delta_p=1.8 \gamma$ shown in Fig.~\ref{fig:shift}(b). For clarity, we consider the three different values of control field $\Omega=2 \gamma$ (solid) $\Omega=3 \gamma$ (dashed), and $\Omega=4 \gamma$ (dotted) in order to make one-to-one correspondence with results of $|r_s / r_p|$ in Fig.~\ref{fig:rsp}. 
It is shown that  $\delta_{p}^{+}$ give extreme values in the vicinity of $\theta_{B}$ where $r_s / r_p$ is large.
The results indicate that with the increase of the control field $\Omega$, the maximum positive and negative spin shift values keep on decreasing and their positions shift almost near the Brewster angle.
This decrease of $\delta_{p}^{+}$ is related to dielectric constant $\epsilon_2$ or susceptibility $\chi$. From the results shown in Fig.~\ref{fig:chi}, the probe susceptibility is purely imaginary at anomalous and normal dispersion regimes and decreases with $\omega$, which results in a decrease in the magnitude of shift~\cite{wan_controlling_2020}.
The transverse shift changes the sign from positive to negative around the Brewster angle $\theta_B$.
The transverse SHE is positive for $\theta_i < \theta_b$, negative for $\theta_i > \theta_b$, and becomes zero at the Brewster’s angle.
One possible reason is that the horizontal component of the probe's electric field alters its phase while the vertical component remains unaltered. 
Therefore, the phase difference $\phi_s - \phi_p$ experiences a $\pi$ phase variation, and the spin accumulation would reverse its directions accordingly. This switching effect at $\delta_{p}=0$ is shown in Fig.~\ref{fig:shift}(c). For $\delta_{p}=1.8 \gamma$ similar switching happened with a less steep shift and was not shown for clarity. 
Due to the reversed spin-dependent splitting, the spin accumulation can be switched by slightly adjusting the incident angle near the Brewster angle.
Enhancing the ratio of $r_s / r_p$ will result in the enhancement of the photonic spin-dependent shift $\delta_{p}^{+}$, which can be effectively controlled by tuning the Rabi frequencies of the control fields as long as the condition of $\Omega \gg \Omega_p$ is satisfied. 
This shows excellent tunability in the spectrum of spin shift by changing the strength of the control field.
For incident angle $\theta_i$, the angular width for the occurrence of the photonic SHE is $\Delta \theta = \theta^{-} - \theta^{+}$. Here, $\theta^{+}$ belong to maximum positive shift while $\theta^{-}$ belongs to maximum negative shift. Therefore, angular width $\Delta \theta$ indicates the range of incident angles at which the Photonic spin Hall shift $\delta_{p}^{+}$ changes sign from positive to negative. 
To clarify the effects of $\Omega$ and probe detuning $\delta_p$, we extracted the $\Delta \theta$ from the results of Fig.~\ref{fig:shift} (a) and (b). 
Figure~\ref{fig:shift} (d) shows that angular width increases with increase of $\Omega$. 
Furthermore, $\Delta \theta$ is larger for the normal dispersion regime (circles) as compared to the anomalous dispersion regime (squares).

Finally, we discuss the spin-dependent shifts in the anomalous dispersion region for $\Omega_1 \neq \Omega_2$. Our numerical results are shown in Fig.~\ref{fig:omega}.
In Fig.~\ref{fig:omega}(a), we first fixed $\Omega_2=2 \gamma$ and show the contour plot of the transverse SHE as a function of $\theta_i$ and $\Omega_1$ at fixed $\delta_p=0$. According to Fig.~\ref{fig:omega}(a), the transverse SHE is positive for $\theta_{i}<\theta_{B}$ and becomes negative for $\theta_{i}>\theta_{B}$. It can be seen that the magnitudes of the positive and negative spin-dependent shifts are larger when $\Omega_1 < \Omega_2$ and the Brewster angle is also slightly shifted to lower incident angles. In Fig.~\ref{fig:omega}(b), we plotted the transverse spin-separation in the normal dispersion regime $\delta_{p}=1.8 \gamma$. For fixed $\Omega_2=2 \gamma$, we show the variation of the photonic SHE with respect to $\theta_i$ and $\Omega_1$ at $\delta_p= \Delta \nu$. Again, one can see that the transverse displacements give extreme values around the Brewster's angle. If we increase the value of $\delta_{p}=1.8 \gamma$, then one can observe that the magnitude of the spin-dependent shift is decreasing. The shift switch sign near the Brewster's angle. Moreover, for $\Omega_1 < \Omega_2$, both the positive and negative shift magnitudes are higher as shown in Fig.~\ref{fig:omega}(b).

\section{Conclusion}
In conclusion, we suggested a model for the control of the photonic Hall shift in reflected light by using a three-level atomic gain medium. The photonic SHE can be enhanced as a result of a large ratio of reflection coefficients for the TE and TM modes in the presence of an atomic gain medium. By adjusting the Rabi frequencies of the control laser fields and probe field detuning, the reflection ratio can be coherently controlled, which significantly enhances photonic SHE. We observed both positive and negative photonic SHE in anomalous and normal dispersion regimes around the Brewster angle. The range of incident angle to switch the sign of photonic SHE can be effectively controlled through the Rabi frequency of the control laser $\Omega$. The control of photonic SHE is based on the tunable complex susceptibility via external parameters, which does not require a change of structure. Since gain-assisted superluminal light propagation was already observed in experiments using atomic cesium vapor cell at $30^{\circ} \mathrm{C}$ \cite{wang_gain-assisted_2000}. Our proposed results may be detected in the experiment by incorporating the weak measurement protocol near the Brewster angle~\cite{luo_enhanced_2011}. The flexible control of the spin-dependent splitting may have potential applications in cavity QED devices. Furthermore, in cavity QED, it would be interesting to explore the photonic SHE for arbitrarily polarized incidence light ~\cite{kim_generalized_2022, kim_spin_2021} and as well as polarization independent photonic SHE~\cite{kim_incident-polarization-independent_2022}.

\section{acknowledgments}
We acknowledge the financial support from the postdoctoral research grant YS304023905 and the NSFC under Grant No. 12174346. We acknowledge fruitful discussion with Dr. M. Irfan.

\renewcommand{\bibname}{References}

\bibliography{refr}
\end{document}